\documentstyle [12pt]{article}

\begin{document}

\begin{center}

{\large \bf A Rotating Vacuum \\
            and  \\
            the Quantum Mach's Principle} \\

\vspace{3cm}
{\large\em  R.D.M. De $Paola$\footnote[1]{e-mail: rpaola@lafex.cbpf.br}
        and N.F. $Svaiter$\footnote[2]{e-mail: nfuxsvai@lafex.cbpf.br}}\\

\vspace{0.6cm}
{\it Centro Brasileiro de Pesquisas F\'{\i}sicas,} \\
\vspace{0.1cm}
{\it Rua Dr. Xavier Sigaud 150, Urca} \\
\vspace{0.1cm}
{\it 22290-180  Rio de Janeiro, RJ -- Brazil.} \\

\vspace{2.0cm}

\begin{abstract}
In this work we consider a quantum analog of Newton's bucket experiment in
a flat spacetime: we take an Unruh-DeWitt detector in interaction with a real 
massless scalar field. 
We calculate the detector's excitation rate when it is 
uniformly rotating around some fixed point and the field is prepared in the 
Minkowski vacuum and also when the detector is inertial and the field is in 
the Trocheries-Takeno vacuum state. These results are compared and the 
relations with a quantum analog of Mach's principle are discussed.

\end{abstract}

\end{center}

\vspace{1cm}

Pacs numbers: 04.20.Cv, 04.62.+v

\newpage

\section{Introduction}

Using the fact that it is possible to define a rotating quantum vacuum 
\cite{vit,rate}, in this paper we study an apparatus device interacting 
with 
a scalar field producing distinct situations that raise questions 
analogous to the ones discussed by Mach in the Newton's bucket experiment. 
We will consider this quantized system as an analog of Newton's bucket
experiment and we trace a discussion parallel to Mach's one in this new setup.
We will be working in a flat spacetime and, therefore, we consider 
an analog
of Mach's principle in the absence of matter and also in the quantum level.
These points will be clarified in the text.
For a historical background of the classic problem of the rotating disc,
see ref. \cite{classical}, and for the problem of the definition of a 
rotating quantum vacuum state, see ref. \cite{vit,rate,letaw}.

Put a bucket with water inside to rotate around its axis; because of the 
rotation, in the equilibrium situation the water will have a parabolic shape. 
It is 
therefore possible for an observer to tell whether or not the bucket is 
rotating: if the water is level, it is not; if it is parabolic, it
is rotating. In this sense rotation is an absolute concept. 
Mach states that
inertia is relative to all other masses in the universe, implying that one 
could equally well mantain the bucket fixed and rotate all the 
universe around 
the bucket axis \cite{rindler}, obtaining the same result: water with 
parabolic shape. 

As pointed out above, the possibility of defining a rotating quantum vacuum
state allows us to shed some light on a related problem. We will consider the 
interaction of an Unruh-DeWitt detector \cite{unruh} with a massless 
hermitian 
Klein-Gordon field, this interaction being responsible for possible 
transitions 
between internal states of the detector. Two main independent results 
are used 
as basis for the subsequent discussion: the first one, as we stressed, 
is that 
it is possible to define a rotating quantum vacuum different from 
the Minkowski vacuum state \cite{vit,rate}, and the
second one, well known in the literature, is that the response 
function of a detector travelling in a generic world-line is given by the 
Fourier transform 
of the positive frequency Wightman two-point function. The idea is 
therefore to 
compare the following physical situations, always 
assuming that initially the detector is in its ground state: 
the rotating 
detector interacting with the field in the Minkowski vacuum 
and the inertial 
detector interacting with the field prepared in the Trocheries-Takeno 
vacuum state. 
In the analogy we will trace between this quantized system and 
the Newton's
bucket experiment, we can think of the Trocheries-Takeno 
vacuum state as the analog of putting 
the whole universe to rotate. We calculate then the detector's 
excitation 
rate, 
that is, the probability per unit detector time that it ends up in 
the excited 
state due to the interaction with the field, in these two situations. 
The fact 
that the detector gets excited in both situations clearly indicates 
that, also 
in the quantum level, an observer can tell whether or not there is a
relative rotation, but the question that concerns us here is if the rates 
in both settings are equal or not. It can be noted that all the above 
discussion touches upon questions analogous to the ones raised by Mach 
regarding Newton's bucket experiment, but in this case in a quantum level 
and in a flat spacetime, that is, in the absence of matter.

Newton defined a family of reference frames, the so called inertial frames. 
But what then determines which frames are inertial? This question led 
Newton to introduce the {\it absolute space} and the inertial frames were 
those in state of uniform motion with respect to this absolute space. The 
natural consequence of this assumption is that inertial forces like 
centrifugal forces must arise when the proper frame of the body is 
accelerated 
with respect to this absolute space. Mach could not accept Newton's absolute 
space and believed that inertial forces arise when the proper frame of the 
body 
is accelerated with respect to the {\it fixed stars}.

``For me only relative motion exists... When a body rotates 
relatively to the fixed stars, centrifugal forces are produced, 
when rotates relatively to some other different body not 
relatively to the fixed stars, no centrifugal forces are produced." 
\cite{wheeler}

As was pointed out by Weinberg \cite{Weinberg}, Mach had replaced Newton's 
absolute acceleration with respect to the absolute space by acceleration 
relatively to all other masses in the universe. In this way, the {\it 
centrifugal forces}, which for Newton are caused by the rotation with respect 
to the {\it absolute space}, is regarded by Mach as truly {\it gravitational 
forces} because they arise from the (relative) motion of all other masses in 
the universe. The natural consequence of this is that inertia is determined 
by the surrounding masses, i.e. ``relativity of inertia" (see ref. 
\cite{wheeler} and the vast literature cited therein). If the proper frame  
of a body does not rotate with respect to the distant stars then no Coriolis 
forces arise. 

Einstein obtained an answer for these questions in some place 
between both, Newton and Mach. The equivalence principle lies 
somewhere between these authors.  If someone gives the total renormalized 
stress-tensor of all non-gravitational fields of the universe, then it 
is possible to find the metric tensor via the Einstein equations. With 
the gravitational potentials we can find the connections and solve 
the geodesic equation, to find an {\it inertial frame} or 
a freely falling frame. In this frame the laws of physics are 
those of special 
relativity. 

The reader may wonder whether Mach's principle is 
valid in general 
relativity. It is well known that general relativity admits non-Machian 
solutions such as for example the G\"odel solution \cite{Godel}, but with the 
fundamental problem that it presents closed timelike curves. 
The anti-Machian behaviour of this model resides
in that there is rotation of the matter relative to the local
inertial frames. There is an improvement of G\"odel's cosmological solution, 
the Ozsv\'ath-Sch\"ucking model \cite{Schucking}, which assumes also a 
non-zero 
cosmological constant. This model admits foliation by a sequence of spacelike 
hypersurfaces but presents the same anti-Machian behaviour i.e., there is a 
rotation of the matter relative to the local inertial reference frame. 
We conclude that Mach's principle is not satisfied in these solutions of 
Einstein's equations.

Today it is still a matter of controversy how to give a precise meaning of 
Mach's principle and whether general relativity includes Mach's principle 
or must be modified in order to be consistent with the principle. 
Nevertheless there is a general agreement that the dragging of 
inertial frames by rotating masses, predicted by General Relativity, 
is a Machian effect. The first 
author that did the calculations of such effects was 
Thirring \cite{Thirring}. Using a weak-limit 
to Einstein's equations this author found that a slowly 
rotating massive shell can drag the inertial frames within it. 
Still studying rotating shells, Brill and Cohen and also Orwig again found 
dragging effects \cite{Orwig}. Foucault's pendulum is useful 
to give an insight of how works the dragging of inertial 
frames \cite{rindler}. 
It is known that exactly on the poles the precession of the plane of 
oscillation of the pendulum reaches its maximum value: to
an observer situated on the pole, the plane of oscillation gives one turn 
every twenty four hours. Following Newton one says that the plane of
oscillation is fixed relative to the absolute space while 
the earth gives one 
turn beneath it. But Mach's followers would sustain that the plane of 
oscillation is completely determined by all masses in the 
universe, including,
of course, the earth. In this way, if the earth were alone in the universe, 
its mass would be the sole determiner of the inertial properties of the 
pendulum and therefore earth's rotation alone would ``tell" the 
pendulum how to
precess, whereas for Newton the pendulum would still oscillate relative to
the absolute space and nothing would change. Therefore, according to Mach,
the earth must be dragging along the inertial frames in its vicinity, however
slight may be this dragging in comparison with the influence of all other 
masses in the universe.

In the experimental territory there are some attemps to shed some light on 
these problems. Does the presence of large nearby masses affect the laws of 
motion? Because if Mach were right then a large mass could 
produce small changes in the inertial forces observed in its vicinity, 
whereas if Newton were right, then no such effect would occur.
Cocconi and Salpeter \cite{Salpeter} pointed out that since there 
are large masses near us it is possible to perform experiments to verify if 
these masses affect the inertia of small bodies in the earth. Hughes, 
Robinson and Beltran-Lopez \cite{Lopez} made an extensive series of 
measurements with the purpose to test Mach's principle. According to Mach's 
principle inertial effects are due to the distribution of matter in the universe; 
in this way there should be present a small anisotropy in these effects due to 
the distribution of masses in our galaxy relatively to us. The $Li^{7}$ nucleus 
in the ground state has a spin $3/2$, so it splits in a 
magnetic field into four energy levels, which should be equally spaced 
{\it if the laws of nuclear physics} are rotation invariant. If inertia 
were anisotropic there would appear spaced resonant lines. The results of the 
experiments went in opposite direction to the Mach principle. Nevertheless 
some authors claim that this kind of experiment does no contradict Mach's 
principle since the nuclear forces should also exhibit an anisotropy and it
should be expected a null result in the experiment.  
There are a lot of attempts to incorporate the Mach's principle in general 
relativity, as for example ref. \cite{Raine}, where the author claims 
that the Minkowski spacetime is Machian. In the present paper we will discuss 
some questions related to these, but in a flat spacetime, using some new 
results  concerning the definition of a rotating quantum vacuum. 

The reader may wonder what is the
meaning of Mach's principle in the absence of matter, or even 
how to formulate 
it in such a case. We should have in mind that Mach's principle, 
as formulated
by him and others, does not take into account the possibility that space can
be filled with fields and its very formulation within modern physics is not
yet clarified. In the framework of Quantum Field Theory it is the fields
which are regarded as the fundamental entities, in such a way that they 
constitute the physical content of space. The quantum fields 
of the elementary 
particles are often considered to occupy the whole of spacetime and, in fact, 
matter can be considered as a small perturbation of the fields. How should, 
then, be formulated Mach's principle allowing fields to pervade all space?
When we think of fields, it imediatly comes to mind the Fock 
space of possible 
states that the field can occupy; as in Classical Physics one 
says, with Mach,
that what determines the inertial properties of bodies is all matter in the 
universe, one can ask: in what sense the choice of a quantum field state 
defines a state of motion? As a particular case, to what extent does the 
choice of a quantum {\it vacuum} state determine inertial frames?
It is not our intention, in the present paper, to work in this direction but
only to present a quantized system (detector and field) in which one can
ask questions similar to Mach's ones regarding Newton's bucket experiment.

The paper is organized in the following way: in section 2 we discuss the 
response function of a Unruh-DeWitt detector traveling in inertial or
rotating world-lines interacting with a massless scalar field prepared in the 
Minkowski or Trocheries-Takeno vacuum states. The outcomes of these different
situations allow us to discuss the validity of a quantum analogous of Mach's 
principle in a flat spacetime. Conclusions are given in section 3. 
(In this paper $\hbar=c=1$.)

\section{Inertial and rotating detector excitation rates}

The fact that it is possible to define a rotating quantum vacuum different
from the Minkowski vacuum was proved in \cite{vit,rate}. 
There, a scalar field 
is quantized in both inertial and rotating frames, the coordinates of which 
are related by the Trocheries-Takeno coordinate transformation (see below).
In ref. \cite{vit}, the low velocity approximation was used but in ref.
\cite{rate} a general solution of the Klein-Gordon equation in the
rotating frame was found and the construction of the Hilbert space of the
states of the field for a rotating observer was achieved.
Using the fact that it is possible to find the exact mode solutions for the 
Klein-Gordon equation in terms of both sets of coordinates, one is capable to 
compare the two quantizations by means of the Bogolubov transformations.
The computation of the Bogolubov coefficient $\beta_{ij}$ between inertial
and rotating modes gives a non-vanishing result \cite{rate}, implying that
for a rotating observer the Minkowski vacuum is seen as a many
Trocheries-Takeno particles state.

Let us call $R^{(r)}_{M}$ the response function per unit proper time 
of the monopole detector travelling in a rotating world-line and interacting 
with the field prepared in the Minkowski vacuum, and let us call 
${\cal R}^{(i)}_{T}=R^{(i)}_{T}-R^{(r)}_{T}$ the normalized (in the sense 
explained below) response function
per unit time of an inertial detector in interaction with the field prepared
in the Trocheries-Takeno vacuum state, which is the vacuum state properly
defined by a rotating observer. Are these two 
quantities equal or not? Having 
in mind the analogy with Newton's bucket experiment, this question is the 
analog of: ''Will the shape of the water be the same, whether we put the 
bucket or the whole universe to rotate, keeping the other still?"

In the following we shall be using the results of refs. \cite{vit,rate}. 
In those works it is assumed that the transformation of coordinates from an 
inertial reference frame to a uniformly rotating one is given by the 
Trocheries-Takeno transformations, for which three assumptions are made: 
(i) the transformation laws constitute a group; (ii) for small velocities we 
must recover the usual linear velocity law ($v=\Omega r$); and (iii) the 
velocity composition law is also in agreement with special relativity. In 
fact, 
the above transformation predicts that the velocity of a point at distance $r$ 
from the axis is given by $v(r)=\tanh(\Omega r)$. The Trocheries-Takeno
coordinate transformations read:
\begin{eqnarray}
\label{tak1}
t &=&t'\cosh\Omega r' - r'\theta' \sinh\Omega r',\\
r &=&r',\\
\theta&=&\theta'\cosh\Omega r' - \frac{t'}{r'}\sinh\Omega r',\\
z&=&z'.
\label{tak2}
\end{eqnarray}
It is then possible to write the line element and also the Klein-Gordon 
equation in the rotating frame in terms of Trocheries-Takeno coordinates. A 
complete set of exact solutions of the Klein-Gordon equation was found 
\cite{rate}, being given by $\{u_{qmk},u^*_{qmk}\}$, where
\begin{eqnarray}
 u_{qmk}(t,r,\theta,z)&=&N_2\,e^{ikz}
 \exp\left[i\left(m\cosh\Omega r+\omega r\sinh\Omega r\right)\,\theta\right]
 \nonumber \\
 &\times&\exp\left[-i\left(\frac{m}{r}\sinh\Omega r+
\omega\cosh\Omega r\right)t\right]J_m(qr),
\label{mode2}
\end{eqnarray}
where $\omega^2=q^2+k^2$ and $N_2$ is a 
normalization factor. $m=0,\pm 1,\pm 2,\pm 3,\dots$,
$0\leq q<\infty$ and $-\infty<k<\infty$. One sees that these modes are
well-behaved throughout the whole manifold.
Making use of the transformations (\ref{tak1}-\ref{tak2}) one can show 
that these modes are of positive frequency by using the criterium of di Sessa
\cite{disessa}, which states that a given mode is of positive frequency if
it vanishes in the limit $(t')\rightarrow -i\infty$, where $t'$ is the
inertial time coordinate. On the other hand the $u^*_j$ are modes of negative
frequency, and the field operator is expanded in terms of these modes as:
\begin{equation}
 \phi(t,r,\theta,z)=\sum_{m}\int dq\,dk\,
 \left(a_{qmk}u_{qmk}(t,r,\theta,z)+
 a^{\dagger}_{qmk}u^*_{qmk}(t,r,\theta,z)\right),
\label{expansion2}
\end{equation}
where the coefficients $a_{qmk}$ and $a^{\dagger}_{qmk}$ are, respectively, 
the annihilation and creation operators of the Trocheries-Takeno 
quanta of the field. The vacuum state defined by the rotating observer is 
thus the Trocheries-Takeno vacuum state $\left|0,T\right>$ and it is given by
\begin{equation}
 a_{qmk}\left|0,T\right>=0,\,\,\,\,\,\,\,\,\forall\,q,m,k.
\label{TTV}
\end{equation}
The many-particle states, as defined by the rotating observer, can 
be obtained 
through successive applications of the creation operators on 
this vacuum state.

Having sketched the canonical quantization of the scalar field in 
the rotating
frame, we now pass to consider the probability of excitation of a detector 
which is moving in a circular path at constant angular velocity $\Omega$
and at a distance $R_0$ from the rotation axis, interacting with the scalar 
field. The initial state of the detector is its ground state and for the 
initial state of the field we will consider the two distinct 
vacuum states: the usual 
Minkowski vacuum state and also the Trocheries-Takeno 
vacuum state. The other situation of interest is to find the 
probability of excitation of a detector which is moving inertially in 
interaction with the field in the 
Trocheires-Takeno vacuum. The interaction with the field may 
cause transitions 
between the energy levels of the detector and if it is 
found, after the interaction, in an excited state, one can say that the 
Unruh-DeWitt detector
measured ''particles" as the spectrum of fluctuations of the field.

As a detector we shall be considering mainly the detector model of 
Unruh-De Witt \cite{unruh}, which is a system with two internal energy 
eigenstates with monopole matrix element between these two states different 
from zero. According to standard theory \cite{sciama,ginzburg,nami}, the 
probability of excitation per unit proper time of such 
a system (modulo the selectivity
of the detector, which does not interest us here), or simply, its excitation 
rate, is given by:
\begin{equation}
 R(E)=\int_{-\infty}^{\infty} d\Delta t\,e^{-iE\Delta t}G^{+}(x,x'),
\label{rate}
\end{equation}
where $\Delta t=t-t'$, $E>0$ is the difference between the excited and ground 
state energies of the detector and $G^{+}(x,x')$ is the 
positive-frequency
Wightman function calculated along the detector's trajectory. 
Let us note that the positive-frequency Wightman function is given by
\begin{equation}
 G^{+}(x,x')=\left<0|\phi(x)\phi(x')|0\right>,
\label{wightman}
\end{equation}
where $\left|0\right>$ is the vacuum state of the field, which can either be
$\left|0,M\right>$ or $\left|0,T\right>$. Let us consider first the second
possibility.

If one splits the field operator in its positive and negative frequency parts 
with respect to the Trocheries-Takeno time coordinate $t$,
as $\phi(x)=\phi^{+}(x)+\phi^{-}(x)$, 
where $\phi^{+}(x)$ contains only annihilation operators and $\phi^{-}(x)$ 
contains only creation operators (see Eq.(\ref{expansion2})),
and also considers $\left|0\right>$ as the Trocheries-Takeno vacuum state,
i.e., $\left|0\right>=\left|0,T\right>$ then, using Eq.(\ref{TTV}), one
finds that:
\begin{equation}
 G_{T}^{+}(x,y)=\sum_i u_i(x) u^*_i(y),
\label{wightman2}
\end{equation}
where the subscript $T$ stands for the Wightman function calculated in the
Trocheries-Takeno vacuum state. Considering now the modes given by 
Eq.(\ref{mode2}) and that we are interested
in the situation where the detector is at rest in the Trocheries-Takeno frame,
i.e., $\theta=$\,constant, $z=$\,constant and $r=R_0=$\,constant, one finds:
\begin{equation}
 G_{T}^{+}(x,y)=\sum_{m=-\infty}^{\infty}\int_{0}^{\infty}dq
 \int_{-\infty}^{\infty}dk\,N_2^2\,
 e^{-i[\frac{m}{R_0}\sinh\Omega R_0+\omega\cosh\Omega R_0]\Delta t}J_m^2(qR_0).
\label{wightman3}
\end{equation}
Putting the above expression in Eq.(\ref{rate}), we find:
\begin{equation}
 R_T^{(r)}(E,R_0)=\sum_{m=-\infty}^{\infty}\int_{0}^{\infty}dq
 \int_{-\infty}^{\infty}dk\,N_2^2\,J_m^2(qR_0)
 \int_{-\infty}^{\infty}d\Delta t\,
 e^{-i[E+\frac{m}{R_0}\sinh\Omega R_0+\omega\cosh\Omega R_0]\Delta t}.
\label{rate2}
\end{equation}
(In the above, the subscript $T$ stands for the Trocheries-Takeno 
vacuum and the
superscript $(r)$ stands for the rotating world-line 
followed by the detector.)
The last integral gives us 
$2\pi\delta\left(E+\frac{m}{R_0}\sinh\Omega R_0+\omega\cosh\Omega R_0\right)$,
for which the argument is non-null only if $m<0$; we can take the summation
index to run for $m=1,2,3,...$, leaving us with 
\begin{equation}
 R_T^{(r)}(E,R_0)=2\pi\sum_{m=1}^{\infty}\int_{0}^{\infty}dq
 \int_{-\infty}^{\infty}dk\,N_2^2\,J_m^2(qR_0)\,
 \delta\left(E-\frac{m}{R_0}\sinh\Omega R_0
 +\omega\cosh\Omega R_0\right).
\label{rate3}
\end{equation}
The above expression predicts excitation for the detector for any $R_0\neq 0$, 
and depends in a
non-trivial way on the position $R_0$ where it is put. Note that we  
arrive at the same confrontation between canonical quantum field theory and 
the detector formalism, which was settled by Letaw and Pfautsch and also
Padmanabhan and Singh \cite{letaw}: how is it possible for the {\it rotating} 
detector to be excited 
in the {\it rotating} vacuum? However a crucial distinction 
exists between our
present analysis and the above-mentioned works: we state, as proved in refs.
\cite{vit,rate}, that the rotating vacuum is not the Minkowski vacuum.
The non-null excitation rate, Eq.(\ref{rate3}), is attributed independently
to the non-staticity of the Trocheries-Takeno metric and also 
to the detector
model considered by us, which is capable to be excited through emission 
processes, and these two independent origins were carefully analysed in
\cite{rate}.

We now discuss the other case of putting the detector
in a rotating trajectory and preparing the scalar field 
in the usual inertial 
vacuum $\left|0,M\right>$. Writing $\left|0,M\right>$ for $\left|0\right>$
in Eq.(\ref{wightman}), it is easy to show that the positive frequency
Wightman function is given by:
\begin{equation}
 G_{M}^{+}(x',y')=\sum_j v_j(x') v^*_j(y'),
\label{wightmanM}
\end{equation}
where $M$ stands for the Minkowski vacuum state. As the rate of excitation
Eq.(\ref{rate}) is given in terms of the detector's proper time, we shall
express Eq.(\ref{wightmanM}) in terms of the rotating coordinates, using 
the inverse of Takeno's transformations. Let us begin with 
$G^{+}_M(x',y')$ written in inertial coordinates, with 
identifications $r'_1=r'_2=R_0$ and $z'_1=z'_2$, as demanded for this case:
\begin{equation}
 G^{+}_M(x',y')=\sum_{m=-\infty}^{\infty}\int_{0}^{\infty}dq
 \int_{-\infty}^{\infty}dk N_1^2 
 e^{-i\omega (t'_1-t'_2)+im(\theta'_1-\theta'_2)}J_m^2(qR_0),
\label{inert}
\end{equation}
in which $N_1$ is the normalization of the inertial modes.
The inverse of Takeno's transformations read
\begin{eqnarray}
\label{tak1b}
t' &=&t\cosh\Omega r + r\theta \sinh\Omega r,\\
r' &=&r,\\
\theta'&=&\theta\cosh\Omega r + \frac{t}{r}\sinh\Omega r,\\
z' &=&z.
\label{tak2b}
\end{eqnarray}
Using the above in Eq.(\ref{inert}) and taking note of the fact that the
detector is at rest in the rotating frame, i.e., $\theta_1=\theta_2$, we
see that in this manner the Minkowski Wightman function is a function of
the difference in proper time $\Delta t=t_1 - t_2$, which allows us to
calculate the rate of excitation of the orbiting detector when the field 
is in the Minkowski vacuum:
\begin{equation}
 R_M^{(r)}(E,R_0)=2\pi\sum_{m=1}^{\infty}\int_{0}^{\infty}dq
 \int_{-\infty}^{\infty}dk\,N_1^2\,J_m^2(qR_0)\,
 \delta\left(E-\frac{m}{R_0}\sinh\Omega R_0
 +\omega\cosh\Omega R_0\right).
\label{rateM}
\end{equation}
The result above is very much like Eq.(\ref{rate3}), with the exception
that now the normalization of the inertial modes $N_1$ appears instead of 
$N_2$. In the context of Newton's bucket experiment, the
situation above is the analog of putting the bucket to rotate relative to
the {\it fixed stars} and notice that the fact that $R_M^{(r)}(E,R_0)\neq 0$
is translated as {\it water with parabolic shape.}

Finally, let us suppose that it is possible to prepare the field in the
Trocheries-Takeno vacuum and the detector is in an inertial 
world-line and let
us calculate the excitation rate in this situation:
\begin{equation}
 R_T^{(i)}(E,R_0)=\int_{-\infty}^{\infty} d\Delta t'\,
 e^{-iE\Delta t'}G_T^{+}(x',y'),
\label{rate5}
\end{equation}
where the superscript $(i)$ stands for the inertial world-line followed by
the detector, $\Delta t'$ is the difference in proper time in the inertial
frame, and $G_T^{+}(x',y')$ is given by Eq.(\ref{wightman2}), but now written
in terms of the inertial coordinates. It is 
not difficult to write $G_T^{+}(x',y')$ in terms of the inertial
coordinates, recalling that now the detector is not at rest in the rotating
frame. We have therefore the result that:
\begin{eqnarray}
 R_T^{(i)}(E,R_0)&=&2\pi\sum_{m=-\infty}^{\infty}\int_{0}^{\infty}dq
 \int_{-\infty}^{\infty}dk\,N_2^2\,J_m^2(qR_0)\nonumber \\
 &\times&
 \delta\left(E-\left(\omega\Omega R_0-\frac{m}{R_0}\right)\sinh(2\Omega R_0)
 -(m\Omega-\omega)\cosh(2\Omega R_0)\right).
\label{rate6}
\end{eqnarray}
In order to study the activity of the Trocheries-Takeno vacuum, we calculated
the rate of excitation of an Unruh-De Witt detector in two different 
situations: when it is put in the orbiting and in the inertial world-lines 
(respectively eqs. (\ref{rate3}) and (\ref{rate6})). Eq.(\ref{rate3}), 
which is
the excitation rate of the rotating detector in the Trocheries-Takeno vacuum,
is different from zero and, as discussed already, is an unexpected result. We
put aside this problem and we regard this non-null result as a kind of 
{\it noise} of the Trocheries-Takeno vacuum. Accordingly, we assume that the
Trocheries-Takeno vacuum will leave this same noise on a 
detector travelling
on different world-lines; in other words, when computing the excitation rate
of a detector on a different state of motion in the presence of the
Trocheries-Takeno vacuum, this noise, Eq.(\ref{rate3}), is 
also present in the calculation. Therefore, we choose to normalize the excitation 
rate of a detector in a given world-line $x^{\mu}(\tau)$ in the presence of the 
Trocheries-Takeno vacuum, that is, $R_T^{(x^{\mu}(\tau))}$, by considering the 
difference between $R_T^{(x^{\mu}(\tau))}$ and the noise Eq.(\ref{rate3}): 
$${\cal R}^{(x^{\mu}(\tau))}_{T}\equiv R^{(x^{\mu}(\tau))}_{T}-R^{(r)}_{T}.$$ 
(According to this redefinition, the rotating detector is no more excited in
the Trocheries-Takeno vacuum.) Proceeding in this way, the normalized 
excitation rate for an inertial detector in interaction with the field in the 
Trocheries-Takeno vacuum is given by:
\begin{eqnarray}
 {\cal R}_T^{(i)}(E,R_0)&\equiv&R^{(i)}_{T}-R^{(r)}_{T}\nonumber \\
 &=&2\pi\sum_{m=-\infty}^{\infty}\int_{0}^{\infty}dq
 \int_{-\infty}^{\infty}dk\,N_2^2\,J_m^2(qR_0)\times\nonumber \\
 &&\left[\delta\left(E-\left(\omega\Omega R_0-\frac{m}{R_0}\right)
 \sinh(2\Omega  R_0)-(m\Omega-\omega)\cosh(2\Omega R_0)\right)
 \right. \nonumber \\
 &&-\left.\delta\left(E-\frac{m}{R_0}\sinh\Omega R_0
 +\omega\cosh\Omega R_0\right)\right].
\label{rateT2}
\end{eqnarray}
It remains to be clarified the meaning of $R_0$ in the excitation above. When
studying the quantization in the rotating frame, one has to choose the 
world-line followed by the rotating observer, and this is 
parametrized by two
quantities: the angular velocity $\Omega$ and the distance $R_0$ from the
rotation axis. The vacuum state which appears in such a 
quantization is thus
also indexed by these parameters and this is the origin of $R_0$ in the above.
A similar dependence of the excitation rate of a detector on a 
geometrical
parameter appears, for instance, in the well-known Unruh-Davies effect: a
uniformly accelerated detector interacting with the field in the Minkowski 
vacuum state will absorb particles in the same way as if it 
were inertial and
interacting with the field in a thermal bath, with a temperature that depends
on the proper acceleration of the detector.

Recalling the analogy with Newton's bucket experiment and with 
Mach's questions about it, this last situation is the analog of putting the 
whole universe to rotate while keeping the bucket still. The fact that 
${\cal R}_T^{(i)}(E,R_0)\neq 0$ is translated 
again as {\it water with parabolic shape}, but note that the fact that 
${\cal R}_T^{(i)}(E,R_0)\neq R_M^{(r)}(E,R_0)$ means that in 
the quantum level
these two are distinct physical situations, although in the 
classical level
it may be that Mach's conjecture concerning Newton's experiment is the right
one.

As eqs.(\ref{rateM}) and (\ref{rateT2}) are not equal we conclude that 
putting the detector in a rotating world-line in interaction with the field in 
the Minkowski vacuum is a situation not equivalent to putting it in an 
inertial 
world-line interacting with the field in the Trocheries-Takeno vacuum state. 
In this way we demonstrated that, regarding the 
Trocheries-Takeno coordinate 
transformations between an inertial and a rotating frame, 
the quantum analog of Mach's principle is not valid in a flat spacetime scenario.

\section{Summary}

In this paper, we study an Unruh-DeWitt detector travelling on different
world-lines interacting with a scalar field ensuing distinct situations 
that raise questions analogous to the ones discussed by Mach in the Newton's 
bucket experiment. The calculations of the last section show us that: 
$R^{(r)}_{M}$ the response function per unit detector proper time 
of the monopole detector travelling in a rotating world-line and interacting 
with the field prepared in the Minkowski vacuum, and  
${\cal R}^{(i)}_{T}=R^{(i)}_{T}-R^{(r)}_{T}$, the normalized response function
per unit time of an inertial detector in interaction with the field prepared
in the Trocheries-Takeno vacuum state, which is the vacuum state properly
defined by a rotating observer, are not equal. We conclude 
that the quantum analog of Mach's principle does not work in a flat spacetime. 

A natural extension of this work is to repeat the calculations for the 
scalar field using the Heisenberg equations of motion \cite{heisenberg} 
instead of using first order perturbation theory, as was done for the case 
of a uniformly accelerated observer. In this case the 
contributions of the 
self-reaction and the vacuum fluctuations of the field can be identified
separately. Another direction is to calculate the renormalized vacuum 
expectation value of the stress-energy tensor $T_{\mu\nu}$ in the 
Trocheries-Takeno vacuum state. It is well-known that the response of the 
detector and $\left<0,T\right|T_{\mu\nu}\left|0,T\right>_{ren}$ 
are independent measures of the vacuum activity. The 
explicit calculation of 
$\left<0,T\right|T_{\mu\nu}\left|0,T\right>_{ren}$ can improve our 
understanding of the physical meaning of the rotating vacuum.
We expect to present these calculations elsewhere.

\section{Appendix}

This appendix is intended only to clarify the normalizations $N_1$ and $N_2$
of the inertial and rotating modes, respectively. The definition of the
Klein-Gordon scalar product between two solutions of the wave equation is
given by:
\begin{equation}
 \left(\phi,\chi\right)=-i\int_{\Sigma} d\Sigma^{\mu} \sqrt{-g}\,
 \left[\phi\,\partial_{\mu}\chi^*-\chi^*\partial_{\mu}\phi\right],
\label{scalarprod}
\end{equation}
with $d\Sigma^{\mu}=n^{\mu}d\Sigma$, where $n^{\mu}$ is a future-oriented
unit vector orthogonal to the spacelike hypersurface $\Sigma$. One finds the
normalization of a given set of modes $\{v_i,v_i^*\}$ by demanding that:
\begin{equation}
 (v_i,v_j)=\delta_{ij}.
\label{norm}
\end{equation}
From the expression for the inertial modes,
\begin{equation}
 v_{q'm'k'}(t',r',\theta',z')=N_1\,e^{ik'z'+im'\theta'}
 e^{-i\omega't'}J_{m'}(q'r'),
\label{mode}
\end{equation}
and using as $\Sigma$ the hypersurface $t'=0$, it is easy to find that:
\begin{equation}
 N_1=[2\pi(2\omega')^{1/2}]^{-1}.
\end{equation}

For the normalization $N_2$ of the rotating modes, Eq.(\ref{mode2}), one 
chooses the hypersurface $t=0$; one finds that:
\begin{eqnarray}
 (u_i,u_j)&=&2\pi |N_2|^2 \int_0^{\infty}rdrJ_m(qr)J_{m'}(q'r)
 \left[\frac{m+m'}{r}\sinh(\Omega r)+(\omega+\omega')\cosh(\Omega r)\right]
 \nonumber \\
 &&\times \left(iB_{m,m'}(r,\omega,\omega')\right)^{-1}\left[\exp\left(2\pi
 iB_{m,m'}(r,\omega,\omega')\right)-1\right],
\end{eqnarray}
where
\begin{equation}
 B_{m,m'}(r,\omega,\omega')=(m-m')\cosh(\Omega r)+
 (\omega-\omega') r\sinh(\Omega r).
\end{equation}
If we call $I$ the integral in $r$ above, we find:
\begin{equation}
 |N_2|^2 I=\frac1{2\pi}\delta_{m,m'}\frac{\delta(q-q')}{q}.
\end{equation}

\section{Acknowledgement}

We would like to thank L.H. Ford and V.A. De Lorenci for 
valuable discussions.
This paper was supported by Conselho Nacional de Desenvolvimento 
Cient\'{\i}fico e Tecnol\'ogico (CNPq) of Brazil.

\end{document}